# The Markovian and Memoryless Properties of Visual System:

# Evidence from Serial Face Processing

Jun-Ming Yu[1,2], Haojiang Ying[1*]

[1.] Department of Psychology, Soochow University, Suzhou, China

[2.] Psychology, School of Social Sciences, Nanyang Technological University, Singapore

*Correspondence:

Dr. Haojiang Ying

Department of Psychology

Soochow University

Suzhou, China

Email: hjying@suda.edu.cn

Number of Figures: 3
Number of Tables: 2

## *Abstract:*

The visual system can be viewed and studied as an information processing system. If so, then the visual system should follow specific fundamental properties: either a memory or a memoryless system. Previous studies in serial dependence in vision found that the perception of the current stimulus is positively determined by the previous one. However, we are not entirely sure whether this phenomenon is a Markov processing. In this study, participants were asked to rate the social characteristics (attractiveness, trustworthiness, and dominance) of a face, either followed by the same characteristic (the one-trait condition) or another one (the two-trait condition) in randomized orders. By doing so, we can directly test the contribution of the previous input and output to the current output and thus study the properties of the system. Using Derivative of Gaussian, Markov Chain and Linear Mixed effect modeling, convergent results suggested that the serial dependence was absent and the memoryless and Markovian properties were violated in the two-trait condition when testing both attractiveness and dominance, but not in the other conditions. Thus, different facets of (presumably) the same computational task may follow asymmetrical system properties. The study also develops serial dependence as an effective technique to reveal the relationships between different computation tasks.

# Introduction

Throughout past decades, researchers have dedicated themselves to elucidating the detailed mechanisms of human visual perception. With the notion that vision is a kind of information processing (e.g., Marr, 1982), many researchers studied vision from a computational view and endeavored to summarize the computational mechanisms of vision. It is increasingly acclaimed that scientific theories should shift from micro-level 'reductionism' investigation of individual phenomena to macro-level investigation of the dynamic of the complex system (Fried, 2022; Krakauer et al., 2017; Lorenz, 1963). Studying from the view of a system rather than an individual phenomenon would unveil the complex nature and the underlying mechanism of the whole system, which is more than a linear combination of the phenomena (May, 1976). Under this notion, some viewed human vision computation as a signal processing system (Dennett, 1986). If this notion is true, then visual processing should follow certain principles and properties of the signal processing system.

Individuals are continuously receiving unstable information from the ever-changing environment. To achieve temporal stability (Webster, 2015; Cicchini, Mikellidou & Burr, 2018) as well as efficient coding (Yu & Ying, 2021), the human visual system develops mechanisms that exploit information from the recent past and integrate it into current processing (Kim, Burr, Cicchini & Alias, 2020; Glasauer & Shi, 2022). One of the typical phenomena is 'serial dependence', where perceptual judgment of the current stimulus is biased positively toward the previous one (Fischer & Whitney, 2014; Cicchini et al., 2018; Kim et al., 2020). It has been found that serial dependence occurs not only in low-level visual processing, such as orientation (Fischer & Whitney, 2014; Fritsche, Spaak & de Lange, 2020) and motion (Cont & Zimmermann, 2021) but also in high-level processing, including face perception in identity (Liberman, Fischer & Whitney, 2014), age (Clifford, Watson, White, 2018), facial attractiveness (Fornaciai & Park, 2018; Kramer & Pustelnik, 2021) and other facial traits (Yu & Ying, 2021). As serial dependence has been well established as a ubiquitous and general visual mechanism, it is crucial to understand at what processing stage serial dependence occurs and improve temporal stability.

Although studies in serial dependency have been able to describe many computational principles of serial perception, the algorithm level of it is still unclear (Marr, 1982; Krakauer et al., 2017). Recently, Markov Chain modeling has been introduced to investigate the mechanism (Yu & Ying, 2021), and it captures some facets of serial dependence well. The Markov Chain is a stochastic process in which current states are dependent on the previous states (Çinlar, 2011), a process resembling serial dependence (Yu & Ying, 2021). When participants' response states are discrete and finite (for example, rating on a 1-7 scale), a transitional probability matrix which captures how response states transit among trials can be generated using Markov Chain modeling. The transition patterns of the response states across time can be investigated by matrix analysis. Researchers have used the Markov Chain modeling to investigate serial dependence in facial trait judgment and found a general mechanism of serial dependence among various facial traits (Yu & Ying, 2021). As the Markov Chain modeling is an effective tool to test serial dependence, it leaves an open question as to whether serial dependence follows the Strong Markov property, which means the current states are dependent only on the previous ones. From the information processing perspective (Çinlar, 2011; Oppenheim et al., 1997), the strong Markov property in serial dependence means that the state (judgment response) of the future trial is dependent only on the state of the current trial but not that of the previous trial.

Disentangling serial dependence by separating the effect of the previous representation input and the perceptual output taps into the algorithm level of visual processing from the information processing perspective. Also, as many researchers claimed that serial dependence might be a fundamental processing principle of vision (Cicchini, Mikellidou, & Burr, 2018; van Bergen & Jehee, 2019), then unveiling the system property of serial dependence would be vital. In the present study, we aim to test the Markov property of serial visual perception in the paradigm of serial dependence. If the serial visual perception does have strong Markov property, then the processing should follow some basic principle of a discrete-time (information processing) system. If a system follows the Markov property, then it should be a memoryless system (Çinlar, 2011). The 'memoryless-ness' does not mean the memory (e.g., working memory) in cognitive science. However, it means that the system's current output only depends on the current input but not the previous or future input (Oppenheim et al., 1997). Therefore, if serial perception is indeed a discrete-time Markov chain, governed by both the Markov and memoryless properties, then the visual perception would be affected by the input of the current stimulus (determined by the physical property of the current stimulus) the output of the previous stimulus (the perception of the previous stimulus). On the other hand, if the visual processing system violates the memoryless property of the system, then the serial perception is not Markovian. Thus, the perception of the current stimulus is also affected by the input of previous stimulus, indicating that the system works with 'memory', and apparently not Markovian.

The multiple facets of face perception make it a good candidate to explore the Markovian property in visual perception. It has been well established that face perception automatically involves multiple facets of face information (Freeman & Ambady, 2011; Vuilleumier & Righart, 2011). The advantage of using face perception to study the Markovian and 'memoryless' properties of serial visual perception is that some dimensions of one face processing task may share the same computational mechanism and neural circuits, while some are not. For example, almost all facial trait judgment, regardless of which facial trait is being judged, rely on visual information transiting from early visual cortex to the core face processing area (e.g., FFA, Kanwisher, McDermott, & Chun, 1997). However, individuals can extract multiple dimensions of facial traits and form different social evaluations of one face (Todorov, Gobbini, Evans, & Haxby, 2007; Sutherland, Oldmeadow, Santos, Towler, Michael Burt, & Young, 2013), such as trustworthiness, dominance, and attractiveness, which can also be fulfilled at different neural circuits from the extended face processing regions (e.g., amygdala; Mende-Siedlecki, Said, & Todorov, 2013). Given that serial dependence has been found in facial trait judgment (Ajina, Pollard, & Bridge, 2020; Yu & Ying, 2021), which indicates that facial trait judgment at a specific time point is not independent but dependent on perception from the recent past. Investigating the transition between different facial trait judgment in the serial dependence framework would help elucidate the Markovian properties of serial perception in high-level vision.

Note that, whether the high-level vision follows the Markovian and 'memoryless' properties is not a trivial mathematical modeling question. Many Artificial Neural Networks (ANN) were designed with Markovian property as the mathematical precondition. For instance, researchers found that the forward-backward algorithm of Hidden Markov Model, a kind of Markov model in which the model states are hidden, is essentially the same as the back propagation algorithm of ANNs (Niles & Silverman, 1990). Therefore, some fundamental elements of the state-of-the-art ANNs are following Markovian property. Moreover, there are many Convolutional Neural Networks combined with Markov Random Fields (e.g., Qu, Bengio, & Tang, 2019), a kind of undirected graph network based on multi-connected Markov chains. The Convolutional Neural Networks have been widely used in image crowd counting (Han, Wan, Yao, & Hou, 2017), hyperspectral image classification (Cao, Zhou, Xu, Meng, Xu, & Paisley, 2018), and image segmentation (Duan, Liu, Jiao, Zhao, & Zhang, 2017). As more and more researchers are using ANNs as computational neural models to better understand human cognition, we need to clarify, at this very moment, whether these ANNs follows the same information processing system properties as human cognition system. If the human vision system does not follow the Markov properties, then using some ANNs as neural models to understand vision would be severely problematic.

It is not easy to design a behavioral experiment which directly tests the contribution of the input and the output of the high-level vision system. Although Wang and colleagues (2019) tried to use the perceptual learning as a tool to study the similarities and differences among basic emotions, it is still worth noting that due to the neural base of perceptual learning (e.g., Gilbert, Sigman, & Crist, 2001; Seitz, 2017) and the complex physical structure of expressions (e.g., Xu et al., 2008), the perceptual learning method may not be the best way to unveil the macro-level system structure of high-level vision. On the other hand, one recent study intermixed the spatial frequency and orientation judgment tasks using different stimuli (gabor and dots; Ceylan, Herzog, & Pascucci, 2021). However, testing two totally different tasks may not directly test the contribution of the previous (retinal) input as it may be irrelative to the current inner representation: the two following tasks are too different from each other, therefore the previous (retinal) input would hardly affect the current output even it is computationally feasible. Facial trait judgment, which shares common but different computational and neural mechanisms, can be a good candidate to answer this question. And of course, all facial trait perception can be conducted on the same face automatically. On the other hand, face trait perception is multi-dimensional (Todorov et al., 2007; Oosterhof & Todorov, 2008; Sutherland et al., 2013). Thus, we can show the participants with a stream of faces (the same stimuli input), and they may form the inner representations of the faces automatically while making different judgments at different trials. Therefore, the facial trait judgment allows us to investigate the different effect of representation input and judgment output, and thus is an effective starting point to explore the Markov property in visual processing.

To separate the input and output of previous stimulus in the frame of serial dependence, we randomly inter-mixed different facial traits the participants need to judge in our experimental design. In the two-trait condition (participants were asked to rate one of the two traits of a face), we can investigate whether the current facial judgment is biased by the previous judgment/rating or by the previous representation in the same dimension with the current judgment task. Specifically, participants can judge the current face in terms of either the same (the one-trait condition) or different (the two-trait condition) facial trait with that of the previous face. By doing so, we can separately investigate the representation input and judgment output by creating one-trait and two-trait condition in the mixed judgment task. In the present study, we used mixed-condition trials, but made advantage of the independent but common characteristics of facial trait judgment and mixed the instructions of different facial trait judgment instead of different stimulus types. In this way, the inconsistency between trials is not driven by stimulus type *per se*; but is driven by different task instructions and inner representations. In other words, in the two-trait condition of the mixed-condition blocks, if the trait judgment of the current face is dependent only on the trait judgment output of the previous face; then these results favor the notion that the memoryless and Markov properties occurs in the transition between the two facial traits, suggesting that they are computed via a shared algorithm. On the contrary, if the trait judgment of the current face is dependent on the representation input of the previous face, instead of the judgment output; than the results suggest that the transition between the two facial trait violates the memoryless and Markov properties, indicating that different algorithms and mathematical properties between processing of the two facial traits. The mixed-condition design in the serial dependence of facial trait judgment allow us to investigate the Markov properties in high-level vision and examine the link between different facial traits at the level of mathematical properties in the framework of information processing.

Out of many different social characteristics from faces (Jones et al., 2021; Yu & Ying, 2021), we selected attractiveness, trustworthiness and dominance in facial trait judgment based on three-dimensional model in face evaluation proposed by Sutherland and his colleagues (2013). These three facial traits are the three main characteristic dimensions of face evaluation and serve significant social functions (Sutherland et al., 2013). Notably, many researchers argued that trustworthiness and dominance, as two dimensions (valence-dominance model), would explain the face evaluations on social dimensions (Oosterhof & Todorov, 2008; Jones et al., 2021). Whether attractiveness can be viewed as a separate dimension (or viewed as on the trustworthiness/valence axis) is still debated. For example, to interpret the relationship between attractiveness and trustworthiness, an event-related potential study investigated and compared the neural processing time course of facial attractiveness and trustworthiness (Calve, Gutierrez-Garcia & Beltran, 2018). They found that 'attractiveness impressions precede and might prime trustworthiness inferences' (Calve et al., 2018). But Sutherland and colleagues (2013) argued that facial attractiveness indeed a separated and important social trait with its evolution and computation importance. For example, it has been suggested that facial attractiveness is related to mate selection, facial trustworthiness shows a person's willingness to help, and facial dominance is a good indicator of a person's ability of carrying out the helpful behaviors (Sutherland et al., 2013). Thus, we still selected the attractiveness in our task, and decided to study its relationship with other two facial traits.

Many cognitive scientists treated facial trait judgment as the same computational question, which involves evaluating social characteristics of face. Following this hypothesis, different facets of face perception (e.g., attractiveness, trustworthiness, and dominance) should be symmetrical in computation, which means that they are assumed to be computed via a shared algorithm. They thus shall follow the same mathematical and system property. Then, the processing of these three facial traits would either be all Markovian, or all non-Markovian. On the contrary, if some facial traits are Markovian and some are not, which means that the mathematical property is found different in different facets of face evaluation , then we can speculate that these faces are asymmetrically computed in neural processing.

In this study, we presented faces in a series and asked participants to rate the faces in terms of its attractiveness, trustworthiness, and dominance either within the same block of trials (three traits for three baseline blocks) or facial trait instruction intermixed with each other (three combinations for three testing blocks). The rationale is that facial trait judgment is automatic and can be formed after minimum exposure to a face (Bar, Neta, Linz ,2006; Willis & Todorov, 2006). Therefore, participants would automatically form the inner representation of the different facets of the social characteristics (perceptual output) of the same face (retinal input). And thus, we can scrutinize the impact of the previous input and output of the current perception in a well-controlled fashion.

# Methods

### Participants

26 participants were recruited in our experiment (15 females and 11 males, mean age = 19.92). All participants are with normal or corrected-to-normal vision. This study is approved by the Ethics Committee at Soochow University. We chose this sample size based on the power analysis (G*Power 3.1). The result suggested we need at least 21 participants to reach a power of .95 based on a previous study testing serial dependence of social trait of faces (Yu & Ying, 2021). Considering the fact that we are using different modelling methods, we tested slightly more participants.

### Stimuli

Adapted from previous studies (Burns, Yang & Ying, 2021, Yu & Ying, 2022), we selected the 45 faces from three face databases with ethnic Chinese faces: the Nanyang Facial Emotional Expression Database (N-FEE; Yap, Chan, & Christopoulos, 2016), the Taiwanese Facial Expression Image Database (TFEID; Chen & Yen, 2007) and the database from Wang, Yao, and Zhou (2015). The faces were cropped with an oval mask leaving only the internal region visible. Faces were presented in black and white with luminance equalized by the SHINE toolbox (Willenbockel et al., 2010).

### Apparatus

The experiment was conducted by a laptop running Matlab R2016a (MathWorks) via Psychtoolbox (Brainard, 1997; Pelli, 1997) with a 13.3-inch LCD monitor (from MacBook Pro, 13-inch 2018; spatial resolution at 2560×1440 pixels, refresh rate at 60 Hz). During the test, participants were asked to sit 20.5 cm away from the monitor (each pixel subtended .025 degree of visual angle).

### *Procedure*

The general procedure of this study was adapted from previous study investigating serial dependence in face perception (Burns et al., 2021). In general, the faces were presented in the center of the screen one at a time with randomized orders. Participants were asked to rate the facial trait one at a time. In each single trial, central fixational cross was presented for 1 s, following 1 s presentation of the face stimulus (Figure 1). After the face stimuli disappeared, the instructions appeared, asking participants to rate the presented face image in terms of the specified facial trait on a 7-point Likert scale (e.g., 1 for least attractive and 7 for most attractive) on the keyboard. Using a 7-point Likert scale set the system to be a discrete-time finite-space system, which ease the difficulty of modeling and interpretation. To minimize the possible response bias, participants were instructed to use three different lines of keys (from the keyboard) to response. Specifically, 'q' (for least attractive) to 'u' (for most attractive) for attractiveness rating (using the 'qwertyu' keys), 'a' (for least trustworthy) to 'j' (for most trustworthy) for trustworthiness (using the 'asdfghj' keys), and 'z' (for least dominant) to 'm' (for most dominant) for dominance (using the 'zxcvbnm' keys). Only after participants finished the rating phase, the next trial would commence. To better control the inter-stimulus interval, the inter trial interval was set to be at least 1s, which means if a participants' reaction time is shorter than 1s, the program would automatically adjust the interval to be at 1s.

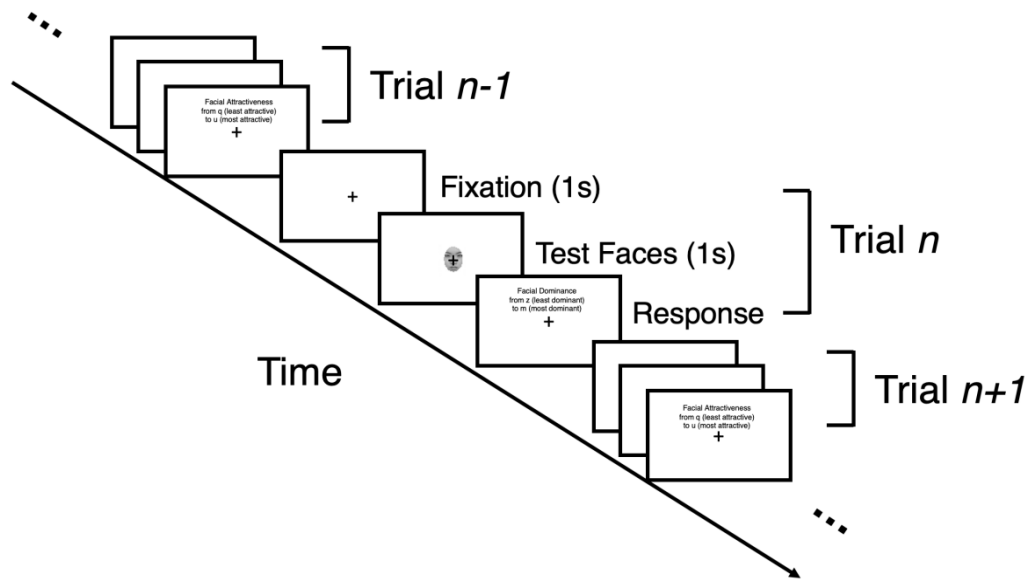

**Figure 1.** The schematic of the experimental procedure. Here is an illustrative example of the 'two-trait' block in which the participants were asked to rate the attractiveness and dominance of each target face in randomized orders. In trial *n-1* and *n+1*, participants were asked to rate the attractiveness of the face, while in trial *n*, they were asked to rate the dominance of a face. In each 'one-trait' block, participants were either asked to rate the face in terms of only one specified facial trait (attractiveness, trustworthiness, or dominance). In each 'two-trait' block, the facial trait judgment instruction was intermixed with each other. Participants need to randomly rate one of two facial trait in randomized orders (attractiveness-trustworthiness, attractiveness-dominance, or trustworthiness-dominance).

There are 6 testing blocks in this experiment: (1) 3 'one-trait' blocks; and (2) 3 'two-trait' blocks. In the one-trait conditions, participants were asked to rate a series of faces in one facial trait (attractiveness, trustworthiness, and dominance) in separated blocks. In the two-trait block, two of the three dimensions (attractiveness-trustworthiness, attractiveness-dominance, trustworthiness-dominance) of facial trait were tested in randomized orders.

To be noted, participants were asked to judge two different facial traits one at a time in the two-trait blocks. For example, in the attractiveness-trustworthiness block, participants could be asked to judge the previous stimulus in terms of its attractiveness, and then the current stimulus in terms of its trustworthiness. In this case, Participant's output of the previous trial from the facial trait they are asked to rate (which is facial attractiveness), but they still receive input of facial trustworthiness as representations for facial traits formed automatically. Therefore, the input and output of the previous trial can be 'two-trait'. It is highly possible that participants' outputs of current trials are influenced by the inputs of the previous trial, instead of the outputs. In other words, serial dependence in high-level visual processing can happened between outputs of the previous trial and inputs of the current trial.

To test this possibility, we further analyzed how the outputs (responses) of the current trials can be influenced by the inputs and the outputs of the previous ones. Here, we considered the output of each trial as the rating from participants, and the input of each trial as the baseline rating of the face stimuli in terms of different facial trait. We used each participant's ratings of face stimuli in the one-trait block to simulate the input values of particular facial trait (differently for each participant) as these two variables highly correlated. To understand serial dependence from the perspective of signal processing, we then use the term 'input' and 'output' in Method and Result section.

We specially selected these three facial traits according to three dimensions of face evaluation proposed by Sutherland and colleagues (2013). In each baseline blocks, participants were asked to rate the faces in terms of only one facial trait. Specifically, if participants were rating in the attractiveness baseline block, they need to rate the face stimuli in terms of facial attractiveness. Participants need to rate the 45 face stimuli twice (90 trials) in each baseline block. In each block, the order of the faces was totally randomized.

The procedure of the three testing blocks was adapted from the baseline block, except that facial trait instructions would change within the testing blocks. Three kinds of facial trait instructions were randomly intermixed, generating three mixed testing blocks (attractiveness-trustworthiness, attractiveness-dominance, trustworthiness-dominance). In the testing block where attractiveness and trustworthiness instructions were intermixed, participants need to judge the presented face stimuli in terms of either attractiveness or trustworthiness (only one trait in a trial). The facial trait being rated was based on the instruction requirements. The 45 face stimuli were rated 2 time for each of the two intermixed facial trait during one testing block. Thus, there are 180 trials in each testing blocks (45 face stimuli × 2 repetitions × 2 facial traits). The orders of the face stimuli were fully randomized across different participants. All participants practiced before formal experiment to ensure they were well familiar with the task.

### *Analysis*

The general logic of the of analysis is: (1) using the Derivative of Gaussian modeling and Markov Chain modeling to validate whether serial dependency occurs in different conditions; (2) using the Derivative of Gaussian modeling to test whether the consistency of the social characteristic judgements (one-trait condition v.s. two-trait condition) affects the serial dependency differently; (3) using the Markov Chain modeling to better model the data from different conditions; and (4) using the Linear Mixed-effect modeling to validate the aforementioned findings.

Before analysis, trials with reaction time smaller than 500ms or larger than 10s were excluded from further analysis (Kramer & Pustelnik, 2021). We used the average rating for each face stimulus in the baseline as the baseline rating for each participant. Three methods were used in the effect of previous faces on the current ones in facial trait evaluation: the conventional Derivative of Gaussian fitting (Fischer & Whitney, 2014), the Markov Chain modeling (Yu & Ying, 2021), and the linear-mixed effect model (Kramer & Jones, 2020; Kramer & Pustelnik, 2021).

The analysis of Derivative of Gaussian fitting and the Markov Chain modeling was conducted with Matlab R2018a (MathWorks) and the linear-mixed effect model was generated based on R (R team) using lme4 package (Bates et al., 2015). We also used Satterthwaite's degrees of freedom estimation method to report significance with lmerTest Package (Kuznetsova et al., 2017). To further investigate how the previous faces' perceptual inputs and rating outputs would influence the current rating outputs, we separated the data from the testing blocks into "one-trait" and "two-trait" inter-trial conditions. According to whether the output trait of the previous and the current trial is one-trait. In the "one-trait" condition, participants rated the face stimuli in terms of the same facial trait between two adjacent trials. In the "two-trait" condition, participants were asked to rate different facial traits for two adjacent trials.

To better test the *memoryless-ness* of the system(s), we employed both the Markov Chain modeling and Linear Mixed effect models. We believe a combination of methods would suit the research question here. The Markov Chain modeling might seem hard to interpret, while the LMM is much direct and easy to interpret. However, the Markov Chain modeling is designed to qualify the Markov property of a system, while the LMM may not unveil such details. For instance, it is very possible that a system (in this case, human vision system) does not utilize the previous input to generate the current output, but there exists a mathematical relationship between these two which can be captured by the LMM. Therefore, combining the two methods would better unveil the system property.

The details of different modeling were described in later sub-sections.

*Derivative of Gaussian fitting*

Serial dependence of facial trait judgment in baseline and testing blocks was first analyzed using the conventional method: the Derivative of Gaussian (DoG) modeling fitting (Liberman et al.,2014; Manassi et al., 2018; Fritsche et al., 2020). In the DoG analysis, we evaluate how the difference between the previous and current ratings would influence participants' ratings of the current face, which indicates the conventional serial dependence effect of previous outputs on the current outputs. For each block, judgment errors were computed as the difference between the participants' ratings and the estimated scores (calculated by averaging the judgments for each face stimuli of each participant in the baseline) of the target face stimuli. Judgment errors were then compared to the differences in ratings between the current and previous responses (Liberman et al., 2014). We pooled the judgment errors of all participants and fitted DoG function.

The DoG was calculated by the function $y = xawce^{-(wx)^2}$, where parameter $x$ is the rating difference between the current and 1-back stimuli (1-back stimuli– current stimuli), $a$ is half the peak-to-trough amplitude of the DoG, $w$ scales the width of the DoG, and $c$ is set to be a constant ($\sqrt{2}/e^{-0.5}$), which scales the curve to make the $a$ parameter equal to the peak amplitude. The amplitude parameter $a$ was taken as the strength of the serial dependence bias, indicating the degree to which participants' judgment of the current stimuli was biased towards the direction of the previous ones (Liberman et al., 2014; Manassi et al., 2017; Fritsche et al., 2020).

Specifically, if participants' judgments were systematically repelled or not influenced by the previous faces, then the parameter $a$ of the DoG should be a negative value or even at zero, respectively. The width parameter $w$ of the DoG curve was treated as a free parameter, constrained to a range of plausible value ($w$ was set between 0 and 5, corresponding to the difference in facial trait value distributed between 0 to 5). We then fitted the Gaussian derivative using constrained nonlinear minimization of the residual sum of squares (Fischer & Whitney, 2014).

### Markov Chain modeling

To further explore the difference in serial dependence between one-trait and two-trait conditions in the testing blocks, the Markov Chain modeling was used to analyze serial dependence. In the present study, participants rated the stimuli on a 1-7 scale. According to our previous study, the Markov Chain modeling is an effective and advantageous tool to capture how participants' rating states transit from trial to trial, and thus well illustrate the amplitude and the transitional pattern of serial dependence with transitional probability matrix (Yu & Ying, 2021).

Specifically, the Markov Chain is a time-series model that expresses transition between finite states according to certain transition probabilities. A Markov process with a state space of {1, 2, …, k} can be determined by the *k*-order transition probability matrix (**A**), which specifies the transition probabilities between any two states. This right-stochastic transition matrix can describe the Markov Chain (i.e., a discrete state-space Markov process). Accordingly, the participants' rating data of the facial trait along the progress of the experiment was regarded as time-series data, and the state space of the Markov model was defined as the rating range {1, 2, …, 7}. The transition information between the rating states was captured by the transition matrix, which contained the probabilities that participants' judgment of the facial trait transited among the seven rating states.

We split the data based on (1) the social characteristic(s) being tested and (2) the consistency of social characteristic(s) of the current and previous trials. We then separately pooled the participants' data together to calculate the transition probability matrices of rating in each condition.

The serial dependence of facial trait judgments could be measured by analyzing the distribution of the transitional probability matrix. More specifically, if ratings in the current trials were biased towards those of the previous trials, which indicated a classic serial dependence, then the transition matrix would be highly likely to be a diagonal matrix (which indicates that the rating of a trial is strictly dependent on a previous trial). If ratings for the facial trait in the current trial were not influenced by a previous trial, then the transition matrix would be very unlikely to correlate with a diagonal matrix. Note that, the Markov Chain modeling is a powerful mathematical method that can summarize the mathematical property of a system having Markovian property. It can also be used for a system that does not satisfy strong Markovian property (Kampen, 1998).

*Linear Mixed-effect Model*

Finally, we used the Linear mixed-effect model (LMM) further to verify the serial dependency in the testing blocks. To analyze the effect of previous rating outputs on the current rating outputs, we carried out the linear mixed-effect model for one-trait and two-trait conditions. In the LMM analysis, we set current rating outputs as the dependent variable. The fixed effects were the intercept, the baseline ratings of the current stimuli, the rating outputs of the previous stimuli, and the random effects were the intercept and the slope of the current stimuli baseline ratings and the intercept of the participant's baseline ratings.

To further investigate how the previous stimuli's perceptual inputs influenced the current rating outputs, we applied the LMM to two-trait inter-trial condition and focused on the effect of the previous inputs influenced the current outputs in the two-trait condition. Specifically, we focus on the previous inputs of the same facial trait with the current outputs in the two-trait condition. We do not list the results of the effect of previous inputs of the same facial trait with the previous output. Because the two variables are highly correlated. As we have investigated the effect of previous outputs on the current inputs, the effect of previous inputs on the current inputs should have similar results. Here, we set previous input as the average baseline rating of the previous face stimuli for each participant. The average baseline ratings were independent of serial dependence, making them more accurate variables to represent previous input in analyzing the two-trait condition of testing blocks.

The reported analysis does not include the effect of inputs in terms of the same facial trait as the previous trait instructions. This is because the values of the previous input were highly correlated with the previous response, making the results highly similar to the effect of serial dependence. As the typical serial dependence has been reported, we would not specially include the effect of previous input in terms of the same facial trait with the previous trait instructions. We did the analysis, and the results show that the effect of previous input, in this case, is highly one-trait with the effect of previous response output. In the previous input analysis, the fixed effects were the intercept, the baseline ratings of the current stimuli, inputs of the previous stimuli, and the random effects were the intercept and the slope of the current stimuli baseline ratings and the intercept of the participant's baseline rating.

# Results

### *Derivative of Gaussian fitting*

Serial dependence was first quantified in the baseline blocks and the testing block separately using the Derivative of Gaussian (DoG) fitting. We plotted the rating error of the current face stimuli as a function of the difference between the previous and the current face rating and then fitted the data into the DoG function. According to the previous study, the amplitude ($\alpha$) of the Derivative of Gaussian is an effective indicator of the size of serial dependence (Fischer & Whitney, 2014). The results showed that the rating errors changed positively as the function of the inter-trial difference in the baseline (for attractiveness, $\alpha = 0.34$; for trustworthiness, $\alpha = 0.45$; for attractiveness, $\alpha = 0.38$), as well as in the three mixed blocks (for attractiveness-trustworthiness, $\alpha = 0.73$; for attractiveness-dominance, $\alpha = 0.61$; for trustworthiness-dominance, $\alpha = 1.08$).

Consistent with the previous study (Liberman et al., 2014), serial dependence was found for facial trait judgment with similar amplitude in the baseline, which demonstrates a general pattern in serial dependence among various facial trait. However, serial dependence among perceptions of different facial traits raised a new question. As previous reports claimed that the serial dependence should be a perceptual bias instead of a response bias (Liberman et al., 2014; Kramer & Pustelnik, 2021), one may wonder whether serial dependence in facial trait judgment arises from the pure perceptual input of the previous face. If this is true, could we further explore the classification of facial trait evaluation and the three-dimensional model in face evaluation in more detail by testing serial dependence among different facial traits (Sutherland et al., 2013)?

To address the issue and effectively analyze the effect of previous inputs and outputs, we separated the serial rating data into two groups based on whether the output attribute (specified facial traits) of the 1-back previous stimulus was consistent with that of the current stimuli. For example, in the attractiveness-trustworthiness testing block, participants were randomly required to judge the presented face stimuli in one of the two facial traits. When the previous facial trait instruction was trustworthiness, and the current was attractiveness, the data was classified as data between two-trait trials. Accordingly, participants were instructed to judge the same facial trait in two successive trials. Data was labeled as in the one-trait condition. In the mixed block, we quantified serial dependence between one-trait trials and between the two-trait trials separately. Results showed that serial dependence exists both in the one-trait condition (for attractiveness-trustworthiness, $\alpha = 0.73$; for attractiveness-dominance, $\alpha = 0.61$; for trustworthiness-dominance, $\alpha = 1.08$) and in the two-trait condition (for attractiveness-trustworthiness, $\alpha = 0.72$; for attractiveness-dominance, $\alpha = 0.75$; for trustworthiness-dominance, $\alpha = 1.17$) for all testing blocks (Figure 2).

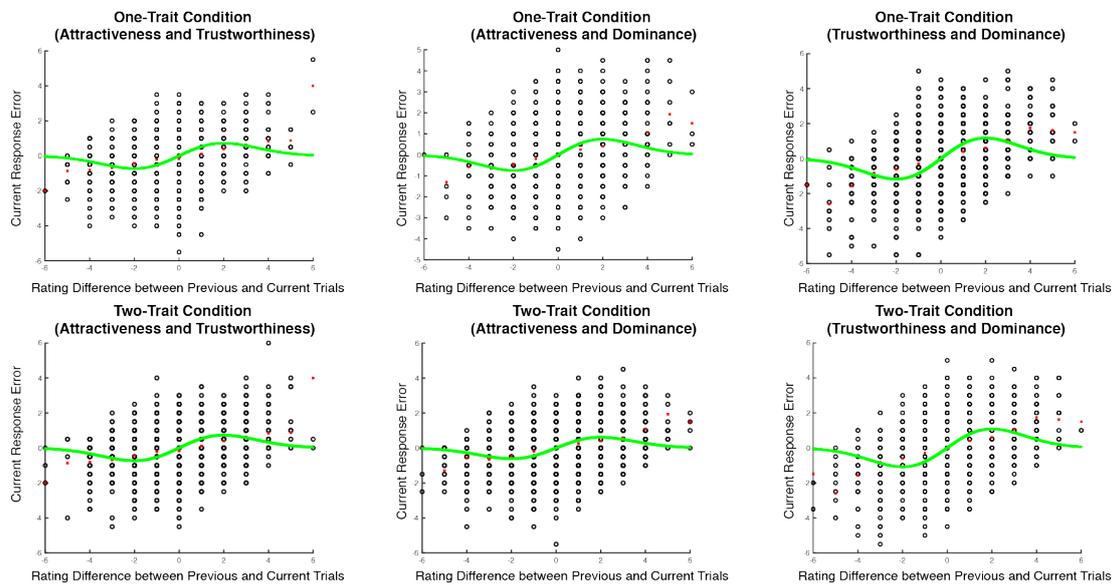

**Figure 2.** Serial dependence and DoG fit between one-trait and two-trait trials in the three testing blocks. The y-axis indicated response errors for the present face, and the x-axis indicated differences in trait rating between the previous and the present face. Each hollow-black dots represent the individual responses. Each solid-red dots represent the average of all participants. The bold-green line is the fitted line of the DoG.

### *Markov Chain Modeling*

We then used the Markov Chain modeling (Yu & Ying, 2021) to explore further the detailed transitional patterns of serial dependence among different trait ratings in the testing blocks (Figure 3). Matrix correlation analysis between the transitional matrices and the diagonal matrix was conducted to quantify the effect of serial dependence. The values of the diagonal direction in the diagonal matrix are one, which represents a pattern of complete transition between the current and the previous rating. The results showed that the correlation between the transitional matrices and the diagonal matrix was significant for all three facial traits in the one-trait condition (all $r$s > .47, all $p$s < .001), which indicates that serial dependence occurred when participants judged the same facial trait between two adjacent trials. However, significant serial dependence was only observed between two-trait trials in the attractiveness-trustworthiness testing block and the trustworthiness-dominance testing block ($r_{att\text{-}tru}$ = 0.37, $p$ < 0.01; $r_{tru\text{-}dom}$ = 0.37, $p$ < 0.01). In the attractiveness-dominance testing block, the correlation between the transitional matrices and the diagonal matrix failed to show serial dependence under the two-trait condition ($r_{att\text{-}dom}$ = 0.10, $p$ = 0.46).

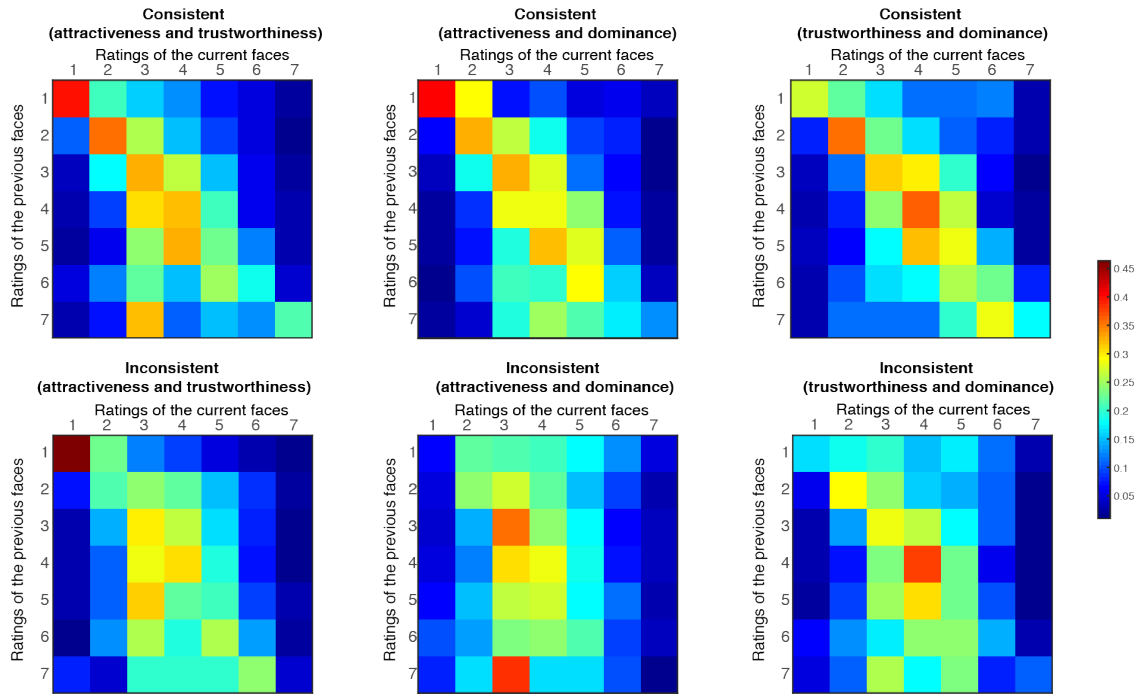

Figure **3.** Transitional probability matrices in the three testing blocks under the one-trait condition and two-trait condition . For each figure, the color of each cell represents the size of transitional probabilities of each current response (y-axis) given a previous response (x-axis). Higher transitional probabilities are marked as red-er, while lower transitional probabilities are marked as blue-er. In the attractiveness-dominance testing block, the correlation between the transitional matrices and the diagonal matrix failed to show serial dependence under the two-trait condition.

### *Linear Mixed-effect Model*

We further analyzed serial dependence in the testing block using the general linear mixed-effect model (LMM; Kramer & Jones, 2020; Kramer & Pustelnik, 2021). As is shown in Table 1, the results of the Linear mixed-effect models summarized the fixed effect of (1) the characteristic of the current (calculated from the baseline rating) and (2) the rating of the previous stimuli (in the previous trial). Also, we sorted the data and conducted the LMM of serial dependence in one-trait and two-trait conditions separately.

Table 1. Parameter estimate for the general linear mixed-effect models in testing the effect of previous ratings in the one-trait and two-trait condition in the mixed blocks.

| | Fixed effect | Estimate | SE | t | p |
|---|---|---|---|---|---|
| ATT&TRU_con | | | | | |
| | Input from current trial | 0.42 | 0.04 | 11.95 | < 0.001 |
| | Output from previous trial | 0.20 | 0.02 | 12.02 | < 0.001 |
| | | | | | |
| ATT&TRU_incon | | | | | |
| | Input from current trial | 0.46 | 0.03 | 15.76 | < 0.001 |
| | Output from previous trial | 0.75 | 0.02 | 4.59 | < 0.001 |
| | | | | | |
| ATT&DOM_con | | | | | |
| | Input from current trial | 0.52 | 0.04 | 14.71 | < 0.001 |
| | Output from previous trial | 0.21 | 0.02 | 12.18 | < 0.001 |
| | | | | | |
| ATT&DOM_incon | | | | | |
| | Input from current trial | 0.55 | 0.04 | 13.30 | < 0.001 |
| | Output from previous trial | -0.01 | 0.02 | -0.20 | **0.84 (n.s.)** |
| | | | | | |
| TRU&DOM_con | | | | | |
| | Input from current trial | 0.38 | 0.06 | 6.33 | < 0.001 |
| | Output from previous trial | 0.20 | 0.02 | 10.84 | < 0.001 |
| | | | | | |
| TRU&DOM_incon | | | | | |
| | Input from current trial | 0.38 | 0.06 | 5.90 | < 0.001 |
| | Output from previous trial | 0.07 | 0.02 | 3.89 | < 0.001 |

The "n.s." stands for not significant.

As expected, the results show that the baseline ratings of the current stimuli positively influence the current rating in all models regardless of inter-trial consistency. Also, serial dependence was found under all the one-trait conditions, as the effect of previous rating outputs was the positive predictor of the current ratings. However, the results turned out to be different in the two-trait conditions. The effects of the previous rating outputs were significant only in two-trait conditions for the attractiveness-trustworthiness block and the trustworthiness-dominance block, but not for the attractiveness-dominance block. In summary, the linear mixed-effect model demonstrates that previous output can positively influence the current output when the specified output attributes (facial trait requirement) are one-trait. However, when it comes to the two-trait condition, the effect of the previous rating output varied across different testing blocks between the three facial traits. Specifically, it failed to show a significant effect of previous rating outputs on the current outputs in the attractiveness-dominance block. However, the phenomenon did happen in the attractiveness-trustworthiness and trustworthiness-dominance blocks.

According to the results mentioned above, the previous rating outputs did not significantly influence the current rating outputs in the two-trait attractiveness-dominance block. However, it does not necessarily mean potential serial dependence did not occur under the interaction between attractive and dominance rating tasks. The previous study investigated serial dependence in face perception as how the current ratings were biased by the previous rating, which defines serial dependence as the effect of rating outputs of the previous trials on the current rating outputs. Nevertheless, as serial dependence has been well established as a perceptual bias (Cicchini et al., 2018; van Bergen & Jehee, 2019), the perceptual inputs of previous face stimuli is also an important factor influencing the current responses. As the perception of facial traits is automatic, when face images appear, participants not only perceive the specific facial trait that is required to be rated but also receive facial trait information in all dimensions. In other words, when the previous facial trait being rated is different from the current facial trait, the inputs of the previous faces from the facial trait as the current facial trait may also influence the current response outputs. To test this probability, we focused on the two-trait condition in the testing blocks and tested how the inputs of the previous faces predict the responses of the current faces.

We analyzed previous input of the same facial trait with the current trial in the two-trait condition. As is mentioned before, the inputs of the previous stimuli were set to be the baseline ratings for each face of each participant, and we matched the previous inputs with the baseline ratings for each two-trait trial in terms of the specific facial traits. Linear mixed-effect models were used to analyze the effect of previous inputs on the current outputs. The fixed effects were the intercept, the baseline ratings of the current stimuli, inputs of the previous stimuli, and the random effects were the intercept and the slope of the current stimuli baseline ratings and the intercept of the participant's baseline ratings (Table 2). The results show that the effect of previous inputs only significantly predicted the current responses in the attractiveness-dominance block. In the attractiveness-trustworthiness and the trustworthiness-dominance block, the effect of previous inputs failed to reach significance.

Table **2.** Parameter estimate for the general linear mixed-effect models in testing the effect of previous input in the one-trait and two-trait condition in the testing blocks.

| | Fixed effect | Estimate | *SE* | *t* | *p* |
|---|---|---|---|---|---|
| ATT&TRU_incon | | | | | |
| | Input from current stimuli | 0.45 | 0.03 | 14.93 | < 0.001 |
| | Input from previous stimuli | 0.03 | 0.02 | 1.53 | **0.13 (n.s.)** |
| | | | | | |
| ATT&DOM_incon | | | | | |
| | Input from current stimuli | 0.54 | 0.04 | 13.36 | < 0.001 |
| | Input from previous stimuli | 0.07 | 0.02 | 3.94 | < 0.001 |
| | | | | | |
| TRU&DOM_incon | | | | | |
| | Input from current stimuli | 0.38 | 0.06 | 5.87 | < 0.001 |
| | Input from previous stimuli | -0.02 | 0.02 | -0.01 | **0.36 (n.s.)** |

The "n.s." stands for not significant.

# Discussion

This study explored the mathematical and system properties of serial face perception and broadened understanding of human vision perception. We tested whether the serial face perception of social characteristics follows Markov property and is a memoryless system or not. To do so, we analyzed serial dependence by investigating whether the outputs of the current trials were exclusively biased by the previous rating outputs (captured by a memoryless process, where current states are dependent only by previous states) or were also influenced by the representation inputs of the previous stimuli. The design of the mixed-condition allowed us to test the Markov property of the transition between different facial traits and revealed the similarities and differences in algorithm and computational mechanism among various facial trait judgments. Using different modeling methods, the results suggested that: (1) serial dependence exists in all conditions but not in the inconsistent condition (the following trial testing a different trait from the previous trial) when testing both attractiveness and dominance in randomized orders; (2) when processing social characteristics, which is widely regarded as the same computational task, the visual system maintained the memoryless property for the attractiveness-trustworthiness block and trustworthiness-dominance block but not for the attractiveness-dominance block.

*The existence of serial dependence in different conditions*

Consistent with previous studies (e.g., Liberman et al., 2014) using other methods and stimuli, serial dependence, in which current rating outputs are positively biased toward the previous rating output, was found for all three facial trait judgments in both the baseline blocks and the mixed testing blocks using DoG fitting and the Markov Chain modeling. In the testing blocks where two facial trait judgments were randomly intermixed, participants judged the previous face stimulus in terms of facial trait, which could either be the same with (one-trait condition) or different from (two-trait condition) the facial trait dimension in the current face.

To examine the serial dependence effect induced by inputs and outputs of previous stimuli separately, we then analyzed serial dependence (biasing effect from previous rating outputs) of the current rating outputs in the one-trait and two-trait conditions. Results from the Markov Chain modeling and the linear mixed-effect model indicate that serial dependence occurred in the one-trait condition for all three testing blocks. However, analysis in the two-trait condition showed that significant serial dependence was only found in the attractiveness-trustworthiness block and the trustworthiness-dominance block but not for the attractiveness-dominance block. As the participants can automatically form a representation of the previous facial trait that is not directly required to be rated in that trial, divergent results from the two-trait condition can directly investigate the effect of previous representation inputs on the current rating outputs. Note that to investigate the biasing effect toward previous representation inputs in the two-trait condition, we especially focused on previous inputs of facial trait that is different from the instructed facial trait in the previous trial. The results showed that rating outputs of the current stimulus are positively biased by the previous rating inputs only in the attractiveness-dominance block but not in the attractiveness-trustworthiness block and the trustworthiness-dominance block. This finding updated our understanding of serial dependence and face perception of social characteristics.

### *The Markov and non-Markov property of serial face perception*

In the present study, the Markov chain modeling and the LMM together studied the 'memoryless-ness' of the visual system. By testing face social characteristics with a tailored design, we investigated the effect of previous outputs and inputs separately on the current inputs in the frame of serial dependence. The memoryless system refers to a system whose current outputs depend only on the current inputs but not previous or future inputs (Oppenheim et al., 1997). Therefore, if the perception output of a face is affected by the input of the current face (the physical property of the current face received from the retina) and the output of the previous face (the perception of the previous face), serial face perception should be a discrete-time Markov process system. However, if the perception of the current faces is also affected by the inputs of the previous faces, then the system is with 'memory' and apparently not Markovian.

The experimental design separated the effect of inputs and outputs (of the system) from the previous trial by randomly asking participants to judge different facial traits in the testing blocks. In this case, when participants judged different facial traits in two successive trials (two-trait condition), we can disentangle the effect of perceptual representation inputs and the rating outputs. Specifically, as participants judged the facial trait of the previous faces, which is different from that of the current faces, they also automatically formed the general perceptual representation of the facial trait evaluation, which includes the same facial trait with the current trials. This general perceptual representation of face evaluation in the previous trial (the previous inputs) may also influence the current rating outputs. If this scenario happens, then the current rating outputs are depend not only on the previous rating output but also on the previous inputs, indicating the visual system is not memoryless in serial face perception. By using the Markov Chain modeling and the linear mixed-effect model in the two-trait condition, we found that the current rating outputs are influenced by the previous rating outputs only in the attractiveness-trustworthiness block and trustworthiness-dominance block but not in the attractiveness-dominance block. Accordingly, the linear mixed-effect model showed that previous inputs significantly influenced the current outputs only in the attractiveness-dominance block, but not in the other two blocks. The convergent finding from Markov Chain modeling and the linear mixed-effect model suggests that the Markov property and memoryless property are not violated in the attractiveness-trustworthiness block and trustworthiness-dominance block but are violated in the attractiveness-dominance block.

### *The understanding of face perception of social characteristics*

By investigating the mathematical properties of the high-level visual system in serial perception, we found different results on the memoryless properties among the three testing blocks, which provide further evidence to unveil the mechanism underlying facial trait evaluation. Sutherland and colleagues (2013) proposed the three-dimensional model of face evaluation, which includes attractiveness, trustworthiness, and dominance. We selected these three basic factors of facial trait judgment and asked participants to randomly rate the face in different facial traits in the mixed testing blocks. Even though significant serial dependence was found for three testing blocks in the one-trait condition, we found that the inputs and outputs of previous stimuli influence current rating output in a different way between the three testing blocks in the two-trait condition. The results from the two-trait condition revealed different mathematical properties in the same processing task and further suggested the common but dissociable computational mechanisms of various facial trait judgment.

The linear mixed-effect model and the Markov Chain modeling suggest that memoryless property was not violated for attractiveness-trustworthiness and dominance-trustworthiness blocks in the two-trait condition. Current rating outputs were positively biased by previous rating outputs but were not influenced by the previous representation inputs in terms of the same facial trait. The results of serial dependence suggest that there may exist overlapping computational mechanisms between attractiveness and trustworthiness, as well as between dominance and trustworthiness, which is in line with the previous study. Dominance and trustworthiness, the two major dimensions/axes of face evaluation, carries significant social signals. One study using the breaking continuous flash suppression found that less dominant and more trustworthy faces are prioritized for awareness (Stein, Awad, Gayet, & Peelen, 2018). The finding indicates that the processing of facial dominance and trustworthiness is involved in extracting social-emotional meanings (Stein et al., 2018). Evidence from neuroimaging also shows that dominance and trustworthiness are processed through similar subcortical pathways involving the amygdala (Ajina et al., 2020). Trustworthiness and attractiveness, two facial trait dimensions closely related to the general valance of face evaluation (Todorov & Engell, 2008), are correlated with each other (Todorov & Engell, 2008). The study also found that there are common neural networking underpinning attractiveness and trustworthiness judgment (such as amygdala, posterior superior temporal sulcus, and medial orbitofrontal cortex, Bzdok et al., 2011; Mende-Siedlecki et al., 2013). More interestingly, the neural time course advantage of attractiveness relative to trustworthiness, which is established by the ERP study, suggests that attractiveness may prime (or bias) trustworthiness judgment (Calve et al., 2018). The priming effect of attractiveness on trustworthiness is consistent with results of our study. Thus, the missing memoryless property in the two-trait condition of the attractiveness-dominance block, which indicates significant effect of previous inputs on the current outputs, leads us to speculate a special relationship between attractiveness and dominance: attractiveness and dominance may has more disparate algorithm and computational mechanisms. The relationships between the three basic trait factors may not be orthogonal but have mutual and complicated relationships. Future study can further investigate the relationship between the three basic factors of facial evaluation from the perspective of visual perception.

The processing of facial trustworthiness, dominance and attractiveness and other social traits in face perception was considered to be the same level (computation) of face perception: the social evaluation of faces (Oosterhof & Todorov, 2008; Jones et al., 2021). Therefore, it is reasonable to assume that these different facets of face evaluation should be symmetrical in computation. Evaluation of different facial traits is presumed to be computed via the same algorithm and thus shall follow the same mathematical property. Then, the processing of these three facial traits would either be all Markovian or all non-Markovian. On the contrary, if some facial traits are Markovian and some are not, which means that the mathematical property is found different in different facets of face evaluation , then we can speculate that these faces are asymmetrical. In the present study, the divergent results on the memoryless property in the facial traits judgments suggest different mathematical properties, indicating separable properties in the level of algorithm and implementation. By exploring the mathematical properties of information processing in the visual system, we can understand vision at the level of the algorithm and provide further insight into the underlying mechanism.

*Mathematical principle and (a)symmetry of cognition*

The present study explored the Markov property in high-level vision, which is not simple mathematical modeling, but inspires the reflection on the current practice of using Artificial Neural Networks as brain/cognitive models. Many Artificial Neural

Networks (ANN) were designed with Markovian property as a hidden presumption. Our results here suggested that certain facets of the same computational task may violate basic properties of Markov property, but some other facets may not. This alerts us that we shall carefully check the mathematical principles of human cognition before fully adapting the ANN models. Future studies should study other cognitive tasks and other mathematical principles of the visual system.

On the other hand, the findings here also highlighted that the same computational task might be symmetrical from a cognitive science perspective but asymmetrical from a mathematical perspective. The contribution of the input from previous stimuli to the current output in the attractive-dominance condition is not only a staggering difference from other social trait pairs but suggests that the mathematical principle is asymmetrical in the presumably the same computational task (the perception of social characteristics). That is, the social characteristics processing may be the same cognitive processing task but not a mathematical computation. Future works should study other asymmetry processing in other aspects of cognitive and the neural system.

# Conclusion

In conclusion, different facets of the same computational task may follow asymmetry mathematical properties of a system. Using Derivative of Gaussian, Markov Chain and Linear Mixed effect modeling, convergent results suggested that the serial dependence was absent and the memoryless and Markovian properties were violated in the two-trait condition when testing both attractiveness and dominance in randomized orders. However, these properties were preserved in other conditions testing trustworthiness, attractiveness, and dominance. Thus, the visual system may adjust the weight of the previous input and output when updating the internal representation of the current stimuli.

# Author's Contribution

**J.M. Yu**: Investigation (coding and testing), Data Analysis, Data Visualization, and Writing (original draft and suggestion). **H. Ying**: Conceptualization, Data Analysis, Data Visualization, Funding & Resources Acquisition, and Writing (original draft and final draft).

# Acknowledgement

**H. Ying** is supported by the National Natural Science Foundation of China (32200850), Natural Science Foundation of Jiangsu Province (BK20200867), and the Entrepreneurship and Innovation Plan of Jiangsu Province. **J.M. Yu** conducted this study under the *Undergraduate Research Advising Project of Soochow University*.